# Preprint: Virtual Reality Based GIS Analysis Platform

Weixi Wang[a,b], Zhihan Lv[c], Xiaoming Li[a,b,c], Weiping Xu[d], Baoyun Zhang[e], Xiaolei Zhang[f]

a. Shenzhen Research Center of Digital City Engineering, Shenzhen, China
b. Key Laboratory of Urban Land Resources Monitoring and Simulation, Ministry of Land and Resources, Shenzhen, China
c. SIAT, Chinese Academy of Science, China
d. Wuhan University, Wuhan, China
e. Jining Institute of Advanced Technology(JIAT), Jining, China
f. Ocean University of China, Qingdao, P. R. China
lvzhihan@gmail.com

**Abstract.** This is the preprint version of our paper on ICONIP2015. The proposed platform supports the integrated VRGIS functions including 3D spatial analysis functions, 3D visualization for spatial process and serves for 3D globe and digital city. The 3D analysis and visualization of the concerned city massive information are conducted in the platform. The amount of information that can be visualized with this platform is overwhelming, and the GIS-based navigational scheme allows to have great flexibility to access the different available data sources.

## 1 Introduction

Virtual Reality Geographical Information System (VRGIS), a combination of geographic information system and virtual reality technology [4] has become a hot topic. By integrating the friendly interactive interface of Virtual Reality System and spatial analysis specialty of Geographical Information System, WebVRGIS [11] [7] based on WebVR [13] is preferred in practical applications, especially by the geography and urban planning. Accordingly, '3-D modes' has been proved as a faster decision making tool with fewer errors [15]. A parallel trend, the utilize of bigdata is becoming a hot research topic rapidly recently [1]. GIS data has several characteristics, such as large scale, diverse predictable and real-time, which falls in the range of definition of Big Data [2]. As a practical tool, most commonly used functions of VRGIS are improved according to practical needs [9]. For our platform, the customer are the employees of the governmental public service or social service agencies. The junior version of our platform is also planed to open the right to use to public. All the presented functions are extracted from the practical customer needs [16] [12] [10] [8].

This research provides a new effective model of three-dimensional spatial information framework and application for urban construction and development directly, which must significantly improve the technical level and efficiency of urban management and emergency response and bring revolutionary changes to the engineering design and construction management field from two-dimensional drawing to three-dimensional collaborative design and construction.

## 2   System Information Process

Under the single-computer environment, the three-dimensional space analytical components have an access to interface access space to analyze the data which will be treated through the uniform data provided by three-dimensional spatial data engine; according to analysis requirement, the access is made to the interface of general spatial analytical components or comprehensive spatial analytical components, and the analysis result can be returned to database or applied into three-dimensional visualization and professional application through the uniform data access interface. Some key technologies are used to tackle the key issues, etc. massive geographical data storage [29] [6] [22] [28] [21] [5] [23] [19] [20] [14] [27] [24] [25] [17] [18] [26]. The information processing process is shown in figure below:

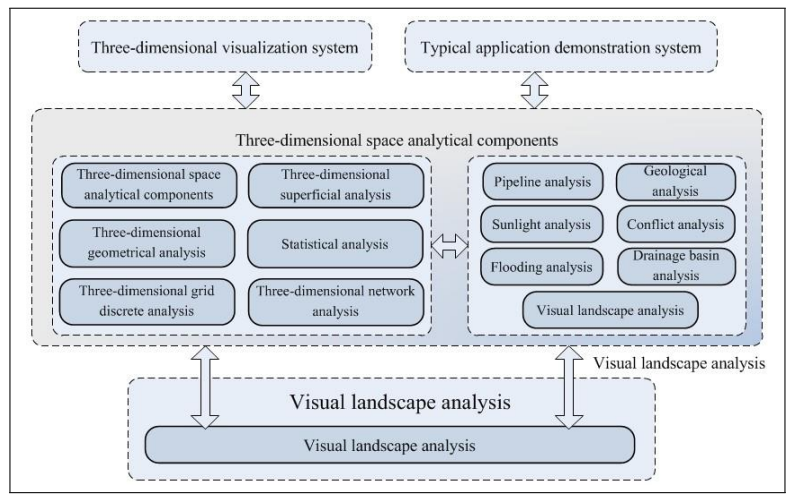

**Fig. 1.** Information process of three-dimensional space analytical components

### 2.1  Three-dimensional Superficial Analytical Module

The surface is an area composed of many points and it contains large quantity of useful information. Through superficial analysis on three-dimensional network, it is able to know the superficial features as a whole, and carry out some specific calculations on existing surfaces, and generate into new data and recognition mode to extract more information. Topographic factor analysis: It mainly includes the calculation of gradient, slope aspect, and curvature. The different topographic factors reflect topographic features from different sides. The gradient and slope aspect are two mutually related parameters: the gradient reflects the inclining degree of slope, and the slope aspect reflects the direction that the slope faces. As the basic elements of topographic feature analysis and visualization, the gradient and slope aspect play a very important role in the research of drainage basin unit, landscape unit, and morphologic measurement. The gradient and slope aspect are used together with other parameters, which can help to solve the application problems such as estimation of forest content, conservation of water and soil, site analysis, and land utilization. Calculation formula of gradient: $\beta = \arctan \sqrt{p^2 + q^2}$ Calculation formula of slope aspect:

$$a = 180° - \arctan \frac{q}{p} + 90° \frac{p}{|p|}$$

, where $p = \frac{\partial H}{\partial x}$, which is the change rate of elevation at $x$ direction; $q = \frac{\partial H}{\partial y}$, which is the change rate of elevation at $y$ direction.

Topographic curvature: It is the reflection of change in shape and concave-convex of topographic curved surface at each section direction, and it is the function of plane point location. The topographic curvature includes plane curvature (contour line curvature) and profile curvature (vertical curvature).

1) The plane curvature refers to the curvature of contour surface (horizontal plane) which passes through ground point and superficial intersection, and it reflects the convergence and divergence mode of superficial physical movement. The calculation formula is shown as below:

$$C_c = -\frac{q^2 r - 2pqs + p^2 t}{(p^2 + q^2)^{\frac{3}{2}}}$$

2) The profile curvature refers to the curvature of normal vector which passes through ground point and the curve which the normal section parallel with the gradient at this point is intersected with topographic curved surface. The profile curvature describes the change in topographic gradient, and it influences the acceleration, slow-down, deposition and flowing state of superficial physical movement. The calculation formula of profile curvature is shown as below:

$$C_p = -\frac{p^2 r + 2pqs + q^2 t}{(p^2 + q^2)(1 + p^2 + q^2)^{\frac{3}{2}}}$$

## 2.2  Three-dimensional Statistical Analysis Module

The spatial correlation analysis mainly focuses on determining the correlation of two or more variables, and the main purpose is to calculate out the degree of correlation and property of relevant variables. The trend-surface analysis is a method in which the spatial distribution and time process of entity features are simulated through mathematical model to predict partial interpolations among actually measured data points of geographic elements under spatial and temporal distribution. Spatial fitting analysis: The so-called fitting refers to the situation that several discrete function values f1,f2,,fn of one function are known, and then several undetermined coefficients f(1, 2,,3) in this function are adjusted to realize minimum difference (significance of least squares) between this function and the known point set. If the undetermined function is linear, this process is called linear fitting or linear regression (in statistics); otherwise this process is called nonlinear fitting or nonlinear regression. The expression can be piecewise function; under this condition, the process is called spine fitting. The spatial interpolation mainly includes Kring and inverse distance weighted method. The Kring interpolation method is one of important contents of spatial statistical analysis method; it is established on the basis of theoretical analysis of semi-variable function, and it is a kind of method of carrying out unbiased optimal estimation for regionalized variable value with finite region. The inverse distance weighted (IDW) is based on similarity principle, that is, the closer two objects are, the more similar their property is; on the contrary, the further two objects are, the less similar their property is.

## 2.3   Three-dimensional Network Analysis Module

The network analysis is one of core problems of GIS spatial analysis function, and the main task and purpose of GIS network analysis function is to carry out geographic analysis and modeling on geographic network (such as traffic network) and urban infrastructure network (such as various kinds of reticle, power lines, telephone lines, and water supply and drainage lines). The network analysis can be used to research and plan how to arrange a network engineering and realize the best operation effect, such as the optimum allocation of certain resource, the shortest operation time or the least consumption from one place to another place.

Network measurement: It is mainly used to measure the incidence relation between the peak and side or the degree of connectivity between peaks in network diagram. The common measurement indexes include: $\beta$ index, number of loops $k$, $a$ index, and $\gamma$ index. As for any three-dimensional network diagram, there exist three kinds of common basic indexes: 1) Number of lines (sides or arcs) ; 2) Number of nodes (peaks) ; 3) Number of sub-graphs in three-dimensional network.

$\beta$ index is also called the rate of line points, and it is the number of average lines for each node in three-dimensional network; the calculation formula is $\beta = \frac{m}{n}$. The loop is a kind of closed path, and its starting point is also the ending point. The number of loops $k$ is the value obtained via subtracting the number of lines $(n - p)$ under minimum-degree connection from the number of actual lines, that is $k = m - n + p$. $a$ index refers to the ratio between the number of actual loops and the possible maximum number of loops in the network. The possible maximum number of loops in the network is obtained via subtracting the number of lines under minimum-degree connection from the possible maximum number of lines. Then, the $a$ index is

$$a = \frac{m - n + p}{\frac{n(n-1)}{2} - (n - 1)} \times 100\%.$$

$\gamma$ index refers to the ratio between the actual number of lines and the possible maximum numbers of lines in the network, and the calculation formula is $\gamma = \frac{m}{n(n-1)/2} \times 100\%$.

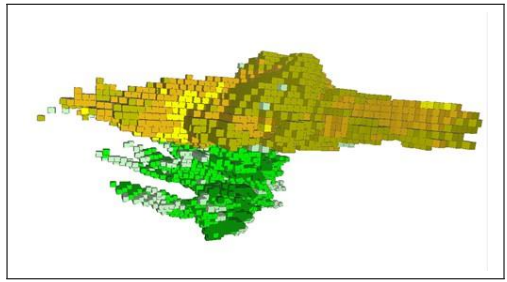

**Fig. 2.** Effect diagram of application of Kring in mine

Optimal path analysis: It refers to realizing the best path selection in three-dimensional network model according to the given parameters. Through establishing three-dimensional path network mode, the users assign starting point and ending point to seek for the nearest path on the network [3].

Indoor and outdoor integrated three-dimensional routing and navigation: A whole process not only includes the route planning of road traffic, but also includes the route planning of indoor architecture. Through inputting the starting point and ending point, the users can analyze the indoor and outdoor integrated three-dimensional route; as for analysis result, it is able to adopt highlighting way to show the found route in the three-dimensional network model.

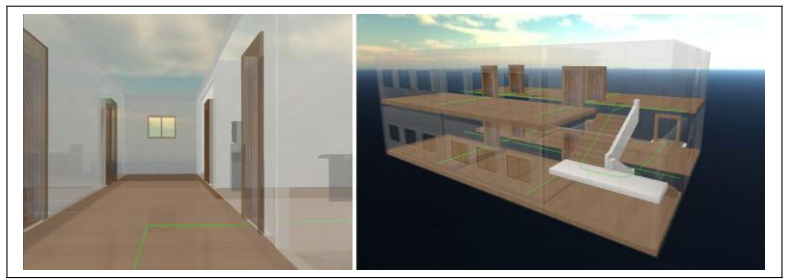

**Fig. 3.** Indoor and outdoor integrated three-dimensional routing and navigation

Connectivity analysis: It refers to analyzing the ability of keeping connection between nodes in the network. Through connectivity analysis on the three-dimensional network model established, the users can obtain the network connectivity and which nodes are adjacent to the node. In this way, it is able to provide network structure data for actual geographic network such as power distribution network, network reconstruction such as pipeline network, state estimation, and safety analysis.

Network address matching: In essence, it refers to inquiry on geographic position and it involves the address coding. The users can enter address list, street network which contains the range of address, and the property value of address to be inquired; through address matching technology, it is able to carry out contrast and matching for the address information entered by users and the address in standard address library, carry out relevance for matched address data, and show it on the map. The network address matching shall be combined with other network analysis functions to meet the complicated analysis required in actual work.

## 3   Analysis on spatial trend surface

The spatial distribution and time process of the solid feature were simulated through the mathematical model of trend surface, and the local interpolation or trend between the measured data point of spatial and temporal distribution of geographical elements was used for analysis.The purpose was to count and analyze the variable trend surface and nature of data set. During use, a group of 3D discrete points were selected as the analysis target, and trend surface was selected in the trend surface analysis dialogue box to analyze the function type and conduct calculation, and in this way, it is possible to give the trend surface analysis coefficient of discrete point set and the trend surface fitness. The figure below is a binary cubic polynomial trend surface generated based on the 3D discrete point coordinate. Aimed at the spatial discrete point set, 3D Nurbs curved surface was used for spatial fitting to analyze the spatial distribution of these discrete points. Nurbs curved surface can be used for the fitting analysis of spatial discrete point set, so as to analyze the spatial distribution of the discrete points. As shown in figure 4, a group of spatial discrete points to be analyzed were selected for spatial fitting analysis, and in this way, it is possible to generate the Nurbs fitting curved surface of these spatial discrete points.

## 4   Conclusion

3D city visualization and analysis platform is a useful tool for the social service agencies and citizens for browsing and analyzing city big data directly, and is agreed upon as being both immediately useful and generally extensible for future applications. The user-need-oriented 3D GIS based smart government portal makes rapid response by real-time and thorough perception of users needs, so as to make timely improvement to the short service board, actively provide convenient, accurate and high-efficiency service to the public and enterprises in online public service.

## Acknowledgments

The authors are thankful to the National Natural Science Fund for the Youth of China (41301439).

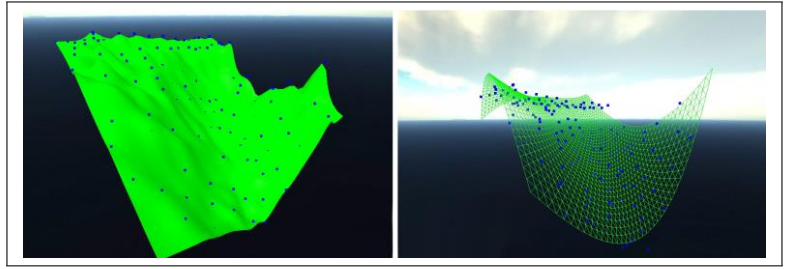

**Fig. 4.** Effect diagram of trend surface analysis